\documentclass[11pt, a4paper]{article}

\usepackage[top=1.5cm, bottom=1.5cm]{geometry}
\usepackage{lscape}

\usepackage[english,francais]{babel}
\usepackage[OT1]{fontenc}
\usepackage[utf8]{inputenc}

\usepackage{url}

\usepackage{hyperref}
\hypersetup{colorlinks=true}

\usepackage{graphicx}
\graphicspath{{figure/}}
\usepackage{epstopdf}

\usepackage{color}
\usepackage{array}
\usepackage{colortbl}
\usepackage{multirow}

\newenvironment{newMargin}[2]%
{\begin{list}{}%
    {\setlength\labelwidth{0pt}%
      \setlength\leftmargin{#1}\item[]%
      \setlength\rightmargin{#2}}\item[]}%
  {\end{list}}

\newcommand{\code}[1]{\texttt{#1}}
\newcommand{\limitations}[1]{\newline {\it Limitations} --- #1}

\newcommand{\libtab}[9]
           {
             \begin{tabular}{|l|l|l|}
               \hline
               #1 & Distribution #2 -- #3 & Doc. updated in #4 \\
               \hline
               \multicolumn{3}{|l|}{\url{#5}}\\
               \hline
               #6 development & #7 developers & First release: #8 \\
               \hline
               \multicolumn{3}{|l|}{License: #9GPL compatible)} \\
               \hline
             \end{tabular}
           }

\newcommand{\parag}[1]{\hspace*{\fill} \newline {\it #1 }}

\newcommand{\libdesc}[8]
{
  {\parag {Types of matrices and vectors available \newline} #1}\\
  {\parag {Linear algebra operations available \newline} #2}\\
  {\parag {Interface with other packages \newline} #3}\\
  {\parag {Performance \newline} #4}\\
  {\parag {Portability: platforms and compilers supported \newline} #5}\\
  {\parag {Installation \newline} #6}\\
  {\parag {Miscellaneous \newline} #7}\\
  {\parag {Limitations \newline} #8}
}

\newcommand{\True}{\cellcolor[gray]{0.1}}
\newcommand{\False}{\cellcolor[gray]{0.95}}

\title{Linear Algebra Libraries}

\author{Claire Mouton \\ \url{claire.mouton@inria.fr}}
\date{March 2009}

\begin{document}
\selectlanguage{english}

\maketitle
\tableofcontents

\newpage
\part{Requirements}
\label{part:requirements}

This document has been written to help in the choice of a linear algebra
library to be included in Verdandi, a scientific library for data
assimilation. The main requirements are
\begin{enumerate}
\item Portability: Verdandi should compile on BSD systems, Linux, MacOS,
  Unix and Windows. Beyond the portability itself, this often ensures that
  most compilers will accept Verdandi. An obvious consequence is that all
  dependencies of Verdandi must be portable, especially the linear algebra
  library.
\item High-level interface: the dependencies should be compatible with the
  building of the high-level interface (e. g. with SWIG, this implies that
  headers (.hxx) have to be separated from sources (.cxx)).
\item License: any dependency must have a license compatible with Verdandi
  licenses (GPL and LGPL).
\item C++ templates, sparse matrices and sparse vectors have to be supported.
\end{enumerate}
Information reported here was collected from December 2008 to March 2009. 

\newpage
\part{CPPLapack}

CPPLapack is a C++ class wrapper for Blas and Lapack.

\libtab {CPPLapack} {beta} {Mar. 2005} {Mar. 2005}
{http://cpplapack.sourceforge.net/} {in} {4} {Apr. 2004} {GPL (}

\libdesc {-- Real double-precision and complex double-precision vectors and
  matrices \\
  -- Real double-precision and complex double-precision band, symmetric and
  sparse matrices}
{-- Eigenvalues computation \\
  -- Linear systems solving \\
  -- SVD decomposition}
{-- Blas \\
  -- Lapack}
{Almost the same as the performance of original Blas and Lapack}
{Platform independent}
{Requires Blas and Lapack}
{-- A few bugs and unsupported Blas and Lapack functions}
{-- Templates not supported, float not supported \\
  -- Alpha version of sparse matrix classes with bugs (the authors advise
  developers not to use these classes in their code) \\
  -- No sparse vectors \\
  -- No separation between headers and sources}

\newpage
\part{Eigen}

Eigen is a C++ template library for linear algebra, part of the KDE project.

\libtab {Eigen} {2.0-beta6} {soon released} {Jan. 2009}
{http://eigen.tuxfamily.org/} {in} {7 contributors, 2} {Dec. 2006} {LGPL and
    GPL (}
\libdesc {-- Dense and sparse matrices and vectors \\
  -- Plain matrices/vectors and abstract expressions \\
  -- Triangular and diagonal matrices \\
  -- Column-major (the default) and row-major matrix storage}
{-- Triangular, SVD, Cholesky, QR and LU solvers \\
  -- Eigen values/vectors solver for non-selfadjoint matrices \\
  -- Hessemberg decomposition \\
  -- Tridiagonal decomposition of a selfadjoint matrix}
{For sparse matrices: TAUCS, umfpack, cholmod and SuperLU}
{Very efficient, see benchmark:
  \url{http://eigen.tuxfamily.org/index.php?title=Benchmark}}
{Standard C++ 98, compatible with any compliant compiler such as \\
  -- GCC, version 3.3 and newer \\
  -- MSVC (Visual Studio), 2005 and newer  \\
  -- ICC, recent versions \\
  -- MinGW, recent versions}
{No dependency}
{-- Templates supported \\
-- Todo includes: interface to Lapack and eigensolver in non-selfadjoint case
  \\
  -- Examples of users: \\
  KDE related projects such as screensavers, kgllib, kglengine2d,
  solidkreator, painting and image editing \\
  Avogadro, an opensource advanced molecular editor \\
  VcgLib, C++ template library for the manipulation and processing of
  triangular and tetrahedral meshes \\
  MeshLab, for the processing and editing of unstructured 3D triangular meshes
  and point cloud \\
  The Yujin Robot company uses Eigen for the navigation and arm control of
  their next generation robots (switched from blitz, ublas and tvmet)}
{-- Sparse matrices and vectors still experimental \\
  -- Eigen 2 is a beta version (Eigen 1 is the old stable version)}

\newpage
\part{Flens}

Flens (Flexible Library for Efficient Numerical Solutions) is a C++ library
for scientific computing providing interface for Blas and Lapack. Flens
intends to be the building block of choice for the creation of serious
scientific software in C++.

\libtab {Flens} {RC1} {Jul. 2007} {Feb. 2008}
{http://flens.sourceforge.net} {in} {9} {2004} {BSD License
  (}
\libdesc {-- General, triangular and symmetric matrix types \\
  -- Storage formats: full storage (store all elements), band
  storage (store only diagonals of a banded matrix), packed storage (store
  only the upper or lower triangular part) \\
  -- Sparse matrix types: general and symmetric, compressed row storage;
  random access for initialization}
{-- Linear systems solving using QR factorization \\
  -- Cg and Pcg methods}
{Blas and Lapack}
{-- Natural mathematical notation: e. g. \code{y += 2 * transpose(A) * x + 1.5
    * b + c}  without sacrificing performances (see section
  \ref{sec:flens_overloaded} in Appendix)\\ 
  -- Very efficient, see benchmarks:
  \url{http://flens.sourceforge.net/session2/tut4.html},
  \url{http://grh.mur.at/misc/sparselib_benchmark/report.html},
    \url{http://flens.sourceforge.net/session1/tut9.html} and section
    \ref{sec:benchmark_seldon}}
{-- Tested on Mac OS X, Ubuntu Linux and a SUSE Opteron cluster \\
  -- GCC: version 4 or higher \\
  -- Intel C++ compiler (icc): version 9.1 \\
  -- Pathscale (pathCC): GNU gcc version 3.3.1, PathScale 2.3.1 driver}
{Requires Blas, Lapack and CBlas}
{-- Extensible: e. g. easy integration of user-defined matrix/vector types \\
  -- Flexible: e. g. generic programming of numerical algorithms \\
  -- Flens implements a view concept for vectors and dense matrices: a
  vector can reference a part of a vector, or also  a row, column or diagonal of a
  matrix; you can apply to these views the same operations as for regular
  vectors and matrices. \\
  --  Templated matrices and vectors with several storage formats}
{-- Lack of portability: not recently tested on Windows (once compiled with
  Microsoft Visual Studio Express compiler with minor modifications in the
  Flens code) \\
  -- No eigenvalues computation \\
  -- No sparse vectors \\
  -- No hermitian matrices}

\newpage
\part{Gmm++}

Gmm++ is a generic matrix template library inspired by MTL and ITL.

\libtab {Gmm++} {3.1} {Sep. 2008} {Sep. 2008}
{http://home.gna.org/getfem/gmm_intro.html} {in} {2 contributors and 2}
{Jun. 2002} {LGPL (}

\libdesc {Sparse, dense and skyline vectors and matrices}
{-- Triangular solver, iterative generic solvers (Cg, BiCgStag, Qmr, Gmres)
  with preconditioners for sparse matrices (diagonal, based on MR iterations,
  ILU, ILUT, ILUTP, ILDLT, ILDLTT) \\
  -- Reference to sub-matrices (with sub-interval, sub-slice or sub-index) for
  any sparse dense or skyline matrix for read or write operations \\
  -- LU and QR factorizations for dense matrices \\
  -- Eigenvalues computation for dense matrices}
{-- Blas, Lapack or Atlas for better performance \\
  -- SuperLU 3.0 (sparse matrix direct solver) for sparse matrices}
{Very efficient, see benchmarks:
  \url{http://grh.mur.at/misc/sparselib_benchmark/report.html} and
  \url{http://eigen.tuxfamily.org/index.php?title=Benchmark}}
{-- Linux/x86 with g++ 3.x and g++ 4.x \\
  -- Intel C++ Compiler 8.0 \\
  -- Linux/Itanium with g++ \\
  -- MacOS X Tiger (with the python and matlab interface) \\
  -- Windows with MinGW and MSys (without the Python and Matlab interface)}
{No special requirement}
{-- Templates supported \\
  -- Examples of users: IceTools, an open source model for glaciers; EChem++:
  a problem solving environment for electrochemistry \\
  -- Gmm++ is included in Getfem++, a generic and efficient C++ library for
  finite element methods, awarded by the second price at the  "Trophées du
  Libre 2007" in the category of scientific softwares \\
  -- Provides a high-level interface to Python and Matlab via Mex-Files for
  Getfem++, covering some functionalities of Gmm++
}
{-- No separation between headers and sources (only header files) \\
  -- No eigenvalues computation for sparse matrices \\
  -- Gmm++ primary aim is not to be a standalone linear algebra library, but
  is more aimed at interoperability between several linear algebra packages.}

\newpage
\part{GNU Scientific Library (GSL)}

GSL is a numerical library for C and C++ programmers. The library provides a
wide range of mathematical routines covering subject areas such as linear
algebra, Blas support and eigensystems.

\libtab {GSL} {GSL-1.12} {Dec. 2008} {Dec. 2008}
{http://www.gnu.org/software/gsl/} {in} {18} {1996} {GPL (}

\libdesc {General vectors and matrices}
{-- Eigenvalues and eigenvectors computation \\
  -- Functions for linear systems solving: LU Decomposition, QR Decomposition,
  SVD Decomposition, Cholesky Decomposition, Tridiagonal Decomposition,
  Hessenberg Decomposition, Bidiagonalization, Householder Transformations,
  Householder solver for linear systems, Tridiagonal Systems, Balancing }
{-- Blas (level 1, 2 and 3)\\
  -- CBlas or Atlas \\
  -- Many extensions such as Marray, NEMO, LUSH (with full interfaces to GSL,
  Lapack, and Blas) and PyGSL}
{Not evaluated}
{-- GNU/Linux with gcc
-- SunOS 4.1.3 and Solaris 2.x (Sparc)
-- Alpha GNU/Linux, gcc
-- HP-UX 9/10/11, PA-RISC, gcc/cc
-- IRIX 6.5, gcc
-- m68k NeXTSTEP, gcc
-- Compaq Alpha Tru64 Unix, gcc
-- FreeBSD, OpenBSD and NetBSD, gcc
-- Cygwin
-- Apple Darwin 5.4
-- Hitachi SR8000 Super Technical Server, cc}
{Easy to compile without any dependencies on other packages}
{}
{-- Can be called from C++ but written in C \\
  -- No sparse matrices and vectors \\
  -- Templates not supported}

\newpage
\part{IT++}

IT++ is a C++ library of mathematical, signal processing and communication
routines. Templated vector and matrix classes are the core of the IT++
library, making its functionality similar to that of MATLAB and GNU
Octave. IT++ makes an extensive use of existing open source libraries
(e. g. Blas, Lapack, ATLAS and FFTW).

\libtab {IT++} {4.0.6} {Oct. 2008} {Oct. 2008}  {http://itpp.sourceforge.net/}
{in} {19 contributors and 11} {2001} {GPL (not L}

\libdesc {-- Diagonal, Jacobsthal, Hadamard and conference matrices \\
  -- Templated vectors and matrices \\
  -- Sparse vectors and matrices}
{-- Matrix decompositions such as eigenvalue, Cholesky, LU, Schur, SVD and
  QR \\
  -- Linear systems solving: over- and underdetermined, LU factorization and
  Cholesky factorization}
{-- Blas, Lapack and FFTW\\
  -- Optionally Atlas, MKL and ACML}
{Not evaluated}
{GNU/Linux, Sun Solaris, Microsoft Windows (with Cygwin, MinGW/MSYS or
  Microsoft Visual C++) and Mac OS X}
{Packages available (Fedora RPM, Debian GNU/Linux and openSUSE)}
{-- Templates supported \\
  -- Separation between headers and sources \\
}
{-- Its main use is in simulation of communication systems and for performing
  research in the area of communications. It is also used in areas such as
  machine learning and pattern recognition.  \\
  -- One important high-performance feature missing in IT++ is the ability to
  create "submatrix-views" and "shallow copies" of matrices, i.e. pass
  submatrices by reference -instead of by value (compared to Lapack++).  }

\newpage
\part{Lapack++}

Lapack++ is a library for high performance linear algebra computations.

\libtab {Lapack++} {beta 2.5.2} {Jul. 2007} {Jul. 2007}
{http://lapackpp.sourceforge.net/} {in} {7} {1993} {LGPL (}

\libdesc {-- Int, long int, real and complex vectors and matrices \\
  -- Symmetric positive definite matrix \\
  -- Symmetric, banded, triangular and tridiagonal matrices}
{-- Linear systems solving for non-symmetric matrices, symmetric positive
  definite systems and solving linear least-square systems; using LU, Cholesky
  and QR matrix factorizations \\
  -- Symmetric eigenvalues computation \\
  -- SVN and QR decompositions}
{-- Blas \\
  -- Lapack}
{High performance linear algebra computation}
{-- Linux/Unix: gcc2.95.x, gcc3.x and gcc4.x \\
  -- Windows 9x/NT/2000: MinGW and gcc3.x \\
  -- Windows 9x/NT/2000: Microsoft Visual Studio, .NET and MSVC \\
  -- Mac OS X}
{Requires Blas, Lapack and a Fortran compiler}
{-- Template functions for matrices}
{-- Templates not supported, float supported only for general matrices \\
  -- No sparse matrices \\
  -- No sparse vectors}

\newpage
\part{Matrix Template Library (MTL)}

MTL is a generic component library developed specially for high
performance numerical linear algebra. MTL includes matrix formats and
functionality equivalent to level 3 Blas.

\libtab {MTL 4} {alpha 1} {Oct. 2007} {Nov. 2008}
{http://www.osl.iu.edu/research/mtl/mtl4/} {in} {4} {1998 (MTL2)} {Copyright
  Indiana University (can be modified to become }
\libdesc {-- Dense2D, morton\_dense and sparse matrices \\
  -- Arbitrary types can be used for matrix elements (float, double, complex)}
{-- Preconditioners: diagonal inversion, incomplete LU factorization without
  fill-in and incomplete Cholesky factorization without fill-in \\
  -- Solvers: triangular, conjugate gradient, BiCg, CgSquared and BiCgStab \\
  -- Iterative methods for solving linear systems thanks to the Iterative
  Template Library (ITL, last release in Oct. 2001): Chebyshev and Richardson
  iterations, generalized conjugate residual, generalized minimal residual and
  (transpose free) quasi-minimal residual without lookahead}
{Blas (optionally Blitz++ thanks to ITL)}
{-- Natural mathematical notation without sacrificing performances: A = B * C
  dispatched to the appropriate Blas algorithm if available; otherwise an
  implementation in C++ is provided (also reasonably fast, usually reaching 60
  percent peak) \\
  -- See benchmarks:
  \url{http://grh.mur.at/misc/sparselib_benchmark/report.html},
  \url{http://eigen.tuxfamily.org/index.php?title=Benchmark} and
  \url{http://projects.opencascade.org/btl/}}
{Can be compiled and used on any target platform with an ANSI C++ compiler \\
  -- Linux: g++ 4.0.1, g++ 4.1.1, g++ 4.1.2, g++ 4.2.0, g++ 4.2.1, g++ 4.2.2,
  icc 9.0 \\
  -- Windows: VC 8.0 from Visual Studio 2005 \\
  -- Macintosh: g++ 4.0.1}
{Requires the Boost library included, optionally scons and a Blas library
  installed}
{-- Templates and generic programming \\
  -- A generic library has been built on top of MTL (ITL:
  \url{http://www.osl.iu.edu/research/itl/})\\
  -- Developed from scratch but inspired by the design and implementation
  details of MTL 2 (interfacing Lapack; supporting sparse, banded,
  packed, diagonal, tridiagonal, triangle, symmetric matrices)}
{-- No sparse vectors \\
  -- No eigenvalues computation \\
  -- No release since the alpha version in Oct. 2007}

\newpage
\part{PETSc}

PETSc, the Portable, Extensible Toolkit for Scientific computation, is a suite
of data structures and routines for the scalable (parallel) solution of
scientific applications modeled by partial differential equations.

\libtab {PETSc} {2.3.3}{May 2007}{Jul. 2008}
{http://www.mcs.anl.gov/petsc/petsc-2/} {in} {11} {Mar. 1999} {Copyright
  University of Chicago (}
\libdesc {-- Parallel vectors and matrices \\
  -- Several sparse matrices storages \\
  -- Symmetric, block diagonal and sequential matrices}
{-- Preconditioners: ILU, LU, Jacobi, block Jacobi, additive Schwartz and ICC
  \\
  -- Direct solvers: LU, Cholesky and QR \\
  -- Krylov subspace methods: GMRES, Chebychev, Richardson, conjugate
  gradients (Cg), CGSquared, BiCgStab, two variants of TFQMR, conjugate
  residuals and Lsqr \\
  -- Nonlinear solvers: Newton-based methods, line search and trust region \\
  -- Parallel timestepping solvers: Euler, Backward Euler and pseudo time
  stepping}
{Blas, Lapack, ADIC/ADIFOR, AnaMod, BlockSolve95, BLOPEX, Chaco, DSCPACK,
  ESSL, Euclid, FFTW, Hypre, Jostle, Mathematica, Matlab, MUMPS, ParMeTiS,
  Party, PaStiX, PLapack, Prometheus, Scotch, SPAI, SPOOLES, SPRNG,
  Sundial/CVODE, SuperLU, Trilinos/ML, UMFPACK}
{Optimal on parallel systems with high per-CPU memory performance}
{Any compiler supporting ANSI C standard on Unix or Windows}
{Requires Blas, Lapack, MPI and optional packages}
{-- Related packages using PETSc such as TAO, SLEPc, Prometheus, OpenFVM,
  OOFEM, DEAL.II and Python bindings (petsc4py and LINEAL) \\
  -- Scientific applications in many fields such as nano-simulations,
  cardiology, imaging and surgery, geosciences, environment,
  computational fluid dynamics, wave propagation and optimization}
{-- No sparse vectors \\
  -- Not coded in C++ but in C \\
  -- Templates not supported, polymorphism used instead}

\newpage
\part{Seldon}

Seldon is a C++ library for linear algebra. Seldon is designed to be efficient
and convenient, which is notably achieved thanks to template classes.

\libtab {Seldon} {2008-11-12} {Nov. 2008} {Nov. 2008}
{http://seldon.sourceforge.net/} {in} {2} {Nov. 2004} {GPL (}
\libdesc {-- Dense and sparse vectors \\
  -- Dense matrices: several formats for rectangular, symmetric, hermitian and
  triangular \\
  -- Two sparse matrix forms: Harwell-Boeing and array of sparse vectors \\
  -- 3D arrays}
{-- Preconditioner of your own or by successive over-relaxations (SOR) \\
  -- Direct dense solvers: LU, QR, LQ and SVD decomposition \\
  -- Iterative solvers: BiCg, BiCgcr, BiCgStab, BiCgStabl, Cg, Cgne, Cgs,
  CoCg, Gcr, Gmres, Lsqr, MinRes, QCgs, Qmr, QmrSym, SYMmetric, Symmlq and
  TfQmr
  \\
  -- Eigenvalues and eigenvectors computation}
{-- Fully interfaced with Blas (level 1, 2 and 3) and Lapack, except for
  functions with banded matrices \\
  -- Direct sparse solvers: MUMPS, SuperLU and UmfPackLU}
{Very efficient, see benchmarks in section \ref{sec:benchmark_seldon}}
{-- Code fully compliant with the C++ standard \\
  -- OS Portable \\
  -- GNU GCC $\geq$ 3 (from 3.2 to 4.3 tested); Intel C++ compiler icc
  (icc 7.1 and 8.0 tested); compile with Microsoft Visual}
{Requires Blas and CBlas}
{-- Thanks to templates, the solvers can be used for any type of matrix and
  preconditioning, not only Seldon matrices: very useful to perform a
  matrix-vector product when the matrix is not stored \\
  -- Provides a Python interface generated by SWIG \\
  -- Exception handling and several debug levels helpful while coding \\
  -- Good coverage of the interface to Blas and Lapack routines
  (see section~\ref{sec:flens_seldon_trilinos_comparisons}) \\
  -- A few alternative functions provided in C++ if Blas is not available \\
}
{-- No band matrices}


\newpage
\part{SparseLib++}

SparseLib++ is a C++ class library for efficient sparse matrix computations
across various computational platforms. 
\begin{newMargin}{0cm}{-2cm}
\libtab {SparseLib++} {v.1.7} {after 1996} {1996}
{http://math.nist.gov/sparselib++/} {minimal maintenance, not in} {3 authors,
  0} {1994} {Public domain (}
 \end{newMargin} 

\libdesc {-- Sparse double vectors \\
  -- Sparse double matrices with several storage formats: compressed
  row/column, compressed diagonal, coordinate formats, jagged diagonal, block
  compressed row and skyline}
{-- Preconditioners: incomplete LU, incomplete Cholesky and diagonal scaling
  \\
  -- Iterative solvers: SOR, Chebyshev iteration, BiCg, BiBgStab, Cg, Cgne,
  Cgne, Cgnr, Gmres, MinRes and Qmr \\
  -- Sparse triangular system solver}
{Blas}
{Not evaluated}
{Various computational platforms}
{Requires Blas}
{-- Built upon sparse Blas (level 3) \\
  -- SparseLib++ matrices can be built out of nearly any C++ matrix/vector
  classes (it is shipped with the MV++ classes by default)} 
{-- Templates not supported (only double elements) \\
  -- No new feature since 1996, only maintenance \\
  -- No eigenvalues computation \\
  -- No dense vectors and matrices}

\newpage
\part{Template Numerical Toolkit (TNT)}

TNT is a collection of interfaces and reference implementations of numerical
objects useful for scientific computing in C++. JAMA/C++ library (Java Matrix
Package translated into C++) utilizes TNT for the lower-level operations to
perform linear algebra operations.

\begin{newMargin}{-0.4cm}{-2cm}
\libtab {TNT} {v.3.0.11} {Jan. 2008} {2003} {http://math.nist.gov/tnt/}
{active maintenance, not in} {2 authors, 0} {Sep. 1999} {Public domain (}
\end{newMargin} 

\libdesc {-- Sparse matrices \\
  -- 1D, 2D and 3D arrays (row-major and column-major)}
{Provided by JAMA/C++: \\
-- SVD decomposition \\
-- SVD, LU, QR and Cholesky solvers \\
-- Eigenvalues computation}
{None}
{Not evaluated}
{ANSI C++ compatibility: should work with most updated C++ compilers (tested
  by the authors with Microsoft Visual C++ v.5.0)}
{By including header files}
{Templates supported}
{-- No sparse vectors \\
  -- No separation between headers and sources (only header files) \\
  -- Beta release since Jan. 2008}

\newpage
\part{Trilinos}

Trilinos provides algorithms and enabling technologies within an
object-oriented software framework for large-scale, complex multi-physics
engineering and scientific problems. Trilinos is a collection of interacting
independent packages (package names are in italic). \\

\libtab {Trilinos} {9.0.1} {Oct. 2008} {Oct. 2008}
{http://trilinos.sandia.gov/} {in} {34} {Sept. 2003} {LGPL (}

\libdesc {-- Core kernel package (\textit{Kokkos}) \\
  -- Dense, symmetric dense, sparse, block sparse, jagged-diagonal sparse
  matrices (\textit{Epetra} and \textit{EpetraExt}) \\
  -- Dense vectors and multivectors (\textit{Epetra} and \textit{EpetraExt}),
  sparse vectors
  (\textit{Tpetra}) \\
  -- Integer and double elements (\textit{Epetra} and \textit{EpetraExt}),
  templates (\textit{Tpetra})}
{-- Preconditioners: ILU-type (\textit{AztecOO} and \textit{IFPACK}),
  Multilevel (\textit{ML}), Block (\textit{Meros}) \\\\
  -- Linear solvers:\\
  Direct dense solvers (\textit{Epetra} and \textit{Teuchos}: wrappers for
  selected Blas and Lapack routines, \textit{Pliris}: LU solver on parallel
  platforms);\\
  Krylov solvers (\textit{AztecOO}: preconditioned Krylov solver,
  \textit{Belos}: iterative solver, \textit{Komplex}: for complex values);\\
  Direct sparse solvers (\textit{Amesos}: for DSCPACK, SuperLU, SuperLUDist
  and UMFPACK);\\
  SVD decomposition (\textit{Epetra}) \\\\
  -- Nonlinear solvers: \\
  System solvers (\textit{NOX}: globalized Newton methods such as line
  search and trust region methods, \textit{LOCA}: computing families of
  solutions and their bifurcations for large-scale applications); \\
  Optimization (\textit{MOOCHO}: reduced-space successive quadratic
  programming (SQP) methods); \\
  Time integration/Differential-algebraic equations (\textit{Rythmos}) \\\\
  -- Eigensolver: block Krylov-Schur, block Davidson and locally-optimal block
  preconditioned conjugate gradient (\textit{Anasazi})}
{-- Uses Blas and Lapack \\
  -- Provides interfaces for Metis/ParMetis, SuperLU, Aztec, Mumps, Umfpack
  and soon PETSc \\
  -- Conjunction possible with SWIG, MPI, Expat (XML parser), METIS and
  ParMETIS}
{\textit{Epetra} provides classes to distribute vectors, matrices and other
  objects on a parallel (or serial) machine}
{Linux, MAC OS X, Windows (under Cygwin), SGI64, DEC and Solaris}
{Requires Blas and Lapack} 
{-- Examples of users: SIERRA (Software Environment for Developing Complex
  Multiphysics Applications), SALINAS (structural dynamics finite element
  analysis), MPSalsa (high resolution 3D simulations with an equal emphasis
  on fluid flow and chemical kinetics modeling), Sundance (finite-element
  solutions of partial differential equations), DAKOTA (Design Analysis Kit
  for Optimization and Terascale Applications~; coding for uncertainty
  quantification, parameter estimation and sensitivity/variance analysis) \\
  -- Most of Trilinos functionalities available in a Python script\\
  -- A package of basic structures with templated types (\textit{Tpetra},
  first release distributed with 9.0.1) \\
  -- \textit{EpetraExt} enables objects to be easily exported to MATLAB \\
  -- Trilinos is based on established algorithms at Sandia. The effort
  required to develop new algorithms and enabling technologies is
  substantially reduced because a common base provides an excellent starting
  point. Furthermore, many applications are standardizing on the Trilinos
  APIs: these applications have access to all Trilinos solver components
  without any unnecessary interface modifications.}
{-- Not yet available: templates (except in an isolated
  package, \textit{Teuchos}) and sparse vectors (Tpetra is still under heavy
  development, release planned in Mar./Apr. 2009)\\
  -- Impossible to build Trilinos under Windows without Cygwin (improved
  Windows support in a further release) \\
  -- Trilinos is a complex collection of interoperable packages and requires
  some careful configuration (with a suitable set of packages and options)}


\newpage
\part{uBlas}

uBlas is a C++ template class library that provides Blas level 1, 2, 3
functionality for dense, packed and sparse matrices.

\libtab {uBlas} {1.33.0} {Jul. 2008} {2008}
{http://www.boost.org/doc/libs/1_35_0/libs/numeric/ublas/} {in} {5}
{2008} {Boost Software License (}

\libdesc {-- Dense, unit and sparse (mapped, compressed or coordinate) vectors
  \\
  -- Dense, identity, triangular, banded, symmetric, hermitian, packed and
  sparse (mapped, compressed or coordinate) matrices}
{-- Submatrices and subvectors operations \\
  -- Triangular solver \\
  -- LU factorization}
{Blas (level 1, 2 and 3)}
{Optimized for large vectors and matrices, see benchmarks:
  \url{http://flens.sourceforge.net/session2/tut4.html},\\
  \url{http://flens.sourceforge.net/session1/tut9.html},
  \url{http://eigen.tuxfamily.org/index.php?title=Benchmark},
  \url{http://projects.opencascade.org/btl/} and 
    and section \ref{sec:benchmark_seldon}}
{OS Independent, requires a modern (ISO standard compliant) compiler such as
  GCC 3.2.3, 3.3.x, 3.4.x, 4.0.x; MSVC 7.1, 8.0; ICC 8.0, 8.1; Visual
  age 6; Codewarrior 9.4, 9.5}
{Requires Blas}
{-- Templates supported \\
  -- Included in Boost C++ libraries \\
  -- Mathematical notation to ease the development (use of operator
  overloading)}
{-- No eigenvalues computation \\
  -- Only basic linear algebra (no linear solving except triangular solver) \\
  -- The implementation assumes a linear memory address model \\
  -- Tuning focussed on dense matrices \\
  -- No separation between headers and sources (only header files)}

\newpage
\part{Other Libraries}


\paragraph{Armadillo++}
is a C++ linear algebra library providing matrices and vectors, interfacing
Lapack and Atlas.  : \url{http://arma.sourceforge.net/}. \limitations {No
  templates (only double), early development (first release in Apr. 2008), no
  portability under Windows without Cygwin.}

\paragraph{Blitz++}
is a C++ class library for scientific computing providing high performance by
using template techniques. The current versions provide dense arrays and
vectors, random number generators, and small vectors and matrices:
\url{http://www.oonumerics.org/blitz/}. \limitations {No sparse matrices and
  vectors. Barely relevant for linear algebra.}

\paragraph{CPPScaLapack}
is a C++ class wrapper for BLACS, PBlas and ScaLapack with MPI. CPPScaLapack
provides a user-friendly interface of high-speed parallel matrix calculation
with Blas and Lapack techonologies for programers concerning with large-scale
computing: \url{http://cppscalapack.sourceforge.net/}. \limitations {Still an
  alpha program, no sparse matrices and vectors, no templates (only
  double-precision vectors and general matrices).}

\paragraph{CVM Class Library}
provides vector and different matrices including square, band, symmetric and
hermitian ones in Euclidean space of real and complex numbers. It utilizes
Blas and Lapack. Contains different algorithms including solving of linear
systems, singular value decomposition, non-symmetric and symmetric eigenvalue
problem (including Cholesky and Bunch-Kaufman factorization), LU
factorization, QR factorization: \url{http://www.cvmlib.com/}. \limitations
{Templates supported but distinction between real and complex types. No sparse
  matrices and vectors.}

\paragraph{IML++}
(Iterative Methods Library) is a C++ templated library for solving linear
systems of equations, capable of dealing with dense, sparse, and distributed
matrices: \url{http://math.nist.gov/iml++/}. \limitations {No matrices and
  vectors, only iterative solvers.}

\paragraph{LA library}
is a C++ interface to Blas and Lapack providing also a general (sparse) matrix
diagonalization, linear solver and exponentiation templates :
\url{http://www.pittnerovi.com/la/}. \limitations {Portability: tested only on
  Linux with a code not fully ANSI C++ compliant.}

\paragraph{LinAl}
The library is based on STL techniques and uses STL containers for the storage
of matrix data furthermore STL algorithms are used where feasible. Low level,
algebraic operators as well as linear solvers and eigenvalue solvers are
implemented, based on calls to Blas, Lapack and CGSOLX and LANCZOS:
\url{http://linal.sourceforge.net/LinAl/Doc/linal.html}. \limitations {No
  vectors, no sparse matrices.}

\paragraph{LinBox}
is a C++ template library for exact, high-performance linear algebra
computation with dense, sparse, and structured matrices over the integers and
over finite fields: \url{http://www.linalg.org/}. \limitations {Does not suit
  to real and complex values.}

\paragraph{MV++}
is a small set of concrete vector and matrix classes specifically designed for
high performance numerical computing, including interfaces to the
computational kernels found in Blas:
\url{http://math.nist.gov/mv++/}. \limitations{Only building blocks to a
  larger-user level library such as SparseLib++ and Lapack++.}

\paragraph{Newmat}
C++ library supports various matrix types. Only one element type (float or
double) is supported.  The library includes Cholesky decomposition, QR,
triangularisation, singular value decomposition and eigenvalues of a symmetric
matrix: \url{http://www.robertnz.net/nm_intro.htm}. \limitations {No sparse
  matrices and vectors, templates not supported.}

\paragraph{RNM}
by \selectlanguage{francais}Frédéric Hecht \selectlanguage{english} provides
C++ classes for arrays with vectorial operations:
\url{http://www.ann.jussieu.fr/~hecht/}. \limitations {Only general matrices,
  no sparse matrices, no vectors, only one linear system solver (conjugate
  gradient), no English documentation.}

\paragraph{SL++ (Scientific Library)}
is a C++ numerical library, composed of modules specialized in various fields
of numerical computations:
\url{http://ldeniau.home.cern.ch/ldeniau/html/sl++.html}. \limitations {Not
  developed since 1998.}

\paragraph{TCBI (Temporary Base Class Idiom)} templated C++ numerical library
implements basic data structures like complex numbers, dynamic vectors, static
vectors, different types of matrices like full matrices, band matrices, sparse
matrices, etc. It also includes some standard matrix solvers like
Gauss-Jordan, LU-decomposition and Singular Value Decomposition and a set of
powerful iterative solvers (Krylov subspace methods along with
preconditioners). Also interfaces to netlib libraries such as CLapack or
SuperLU. Its specificity is being exclusively written in C++, without needing
to interface to Fortran code. The usual loss of performance associated with
object-oriented languages has been avoided through not as obvious
implementations of numerical base classes, avoiding unnecessary copying of
objects. It can be thought of as some sort of reference counting done by the
compiler at compile time. Supported on Linux/Unix with the GNU compiler, on
Windows with the Microsoft Visual C++ (6, 7) compiler and with the Intel
compiler: \url{http://plasimo.phys.tue.nl/TBCI/}. \limitations {No interface
  to Blas and Lapack.}

\newpage
\part{Links and Benchmarks}

\section{Links}

Here are some Web portals related to numerical analysis software and linear
algebra libraries:

-- List of numerical analysis software (Wikipedia)
\url{http://en.wikipedia.org/wiki/List_of_numerical_analysis_software}

-- Numerical computing resources on the Internet (Indiana University)
\url{http://www.indiana.edu/~statmath/bysubject/numerics.html}

-- Scientific computing in object-oriented languages (community resources)
\url{http://www.oonumerics.org/oon/}

-- Scientific computing software (Master's Degrees in Applied Mathematics at
\'Ecole Centrale Paris)
\url{http://norma.mas.ecp.fr/wikimas/ScientificComputingSoftware} \\

\section{Benchmarks}

\subsection{Benchmarks for Linear Algebra Libraries}
~

-- Freely available software for linear algebra on the web -- 2006 comparative
statement \url{http://www.netlib.org/utk/people/JackDongarra/la-sw.html}

-- Benchmark sparse matrices (tests for residual and random order
initialization) -- 2008 \url{http://flens.sourceforge.net/session2/tut4.html}

-- Some Blas Benchmarks -- 2007
\url{http://flens.sourceforge.net/session1/tut9.html}

-- Benchmark 2008 \url{http://eigen.tuxfamily.org/index.php?title=Benchmark}

-- Benchmark of C++ Libraries for Sparse Matrix Computation -- 2007
\url{http://grh.mur.at/misc/sparselib_benchmark/report.html}

-- Benchmark for Templated Libraries project -- 2003
\url{http://projects.opencascade.org/btl/} \\

\newpage
\subsection{Benchmarks including Seldon}
\label{sec:benchmark_seldon}
Platform: Intel Core 2 Duo CPU P9500, 2.53GHz, 6 MB cache, 4 GB
Ram.\\
Compiler: gcc version 4.3.2 (Ubuntu 4.3.2-1ubuntu12).\\
Date: March 2009.\\

\subsubsection {Benchmarks for Dense Matrix}
\label{sec:benchmark_dense}
Adapted from \url{http://flens.sourceforge.net/session1/tut9.html}

\begin{figure}[htpb]
  \centering
  \includegraphics[width=\textwidth]{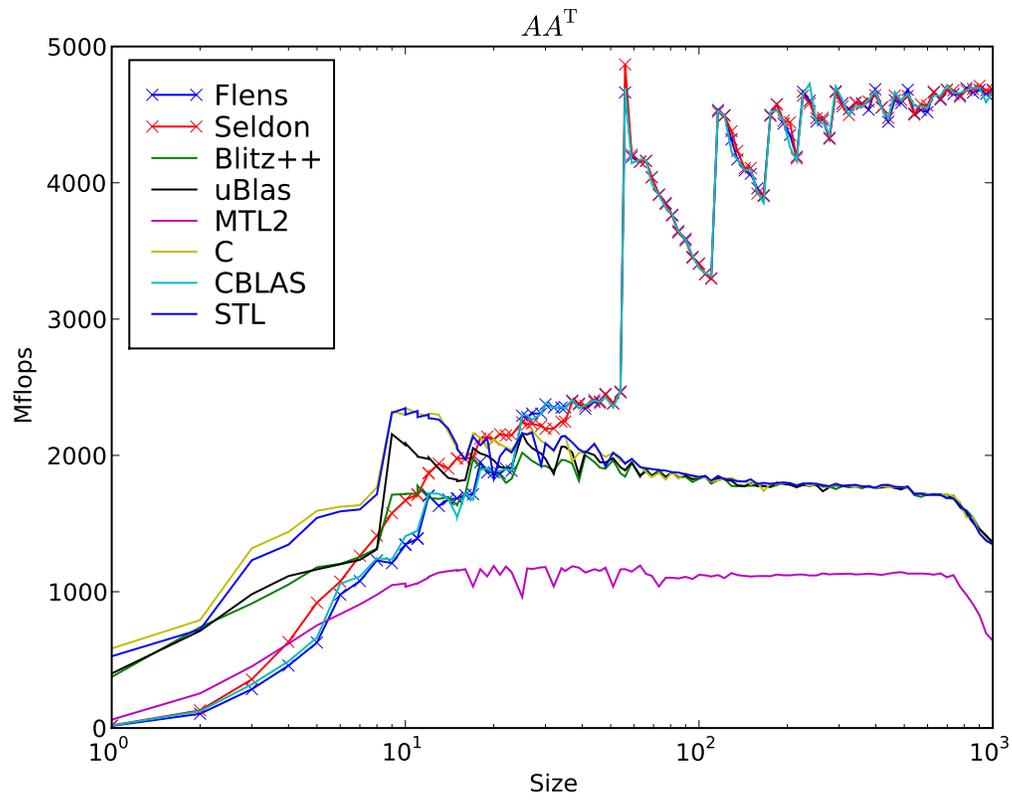}
  \caption{A x At product for dense matrices.}
  \label{fig:dense_aat_product}
\end{figure}

\begin{figure}[htpb]
  \centering
  \includegraphics[width=\textwidth]{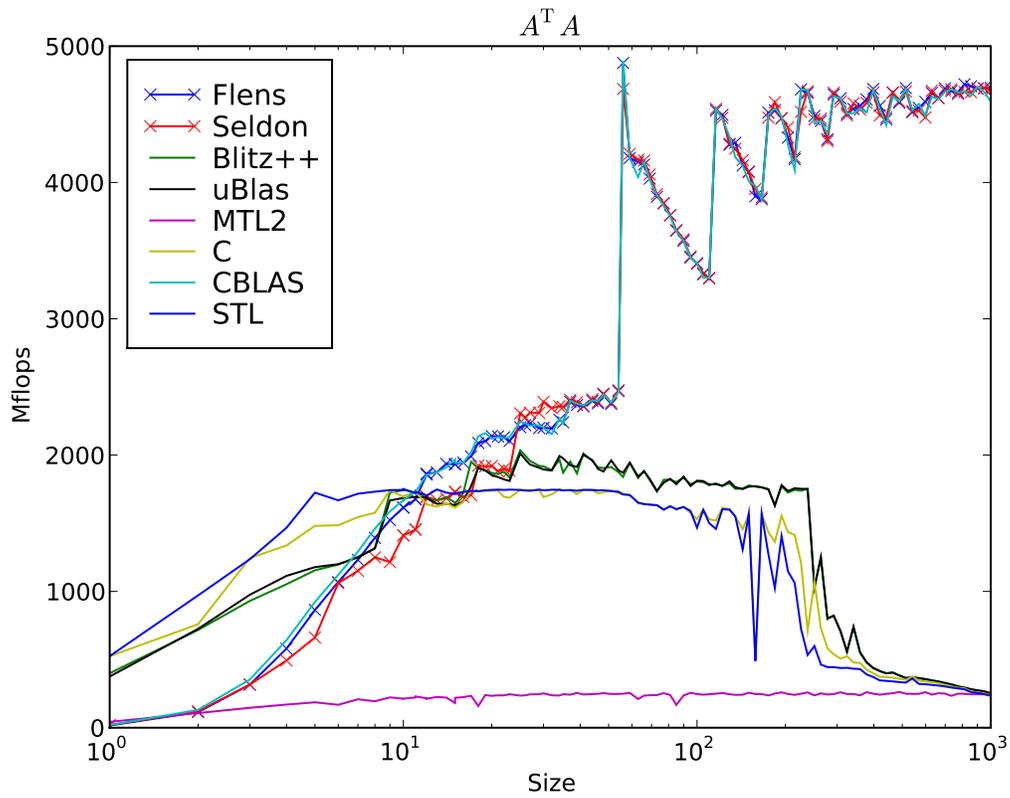}
  \caption{At x A product for dense matrices.}
  \label{fig:dense_ata_product}
\end{figure}

\begin{figure}[htpb]
  \centering
  \includegraphics[width=\textwidth]{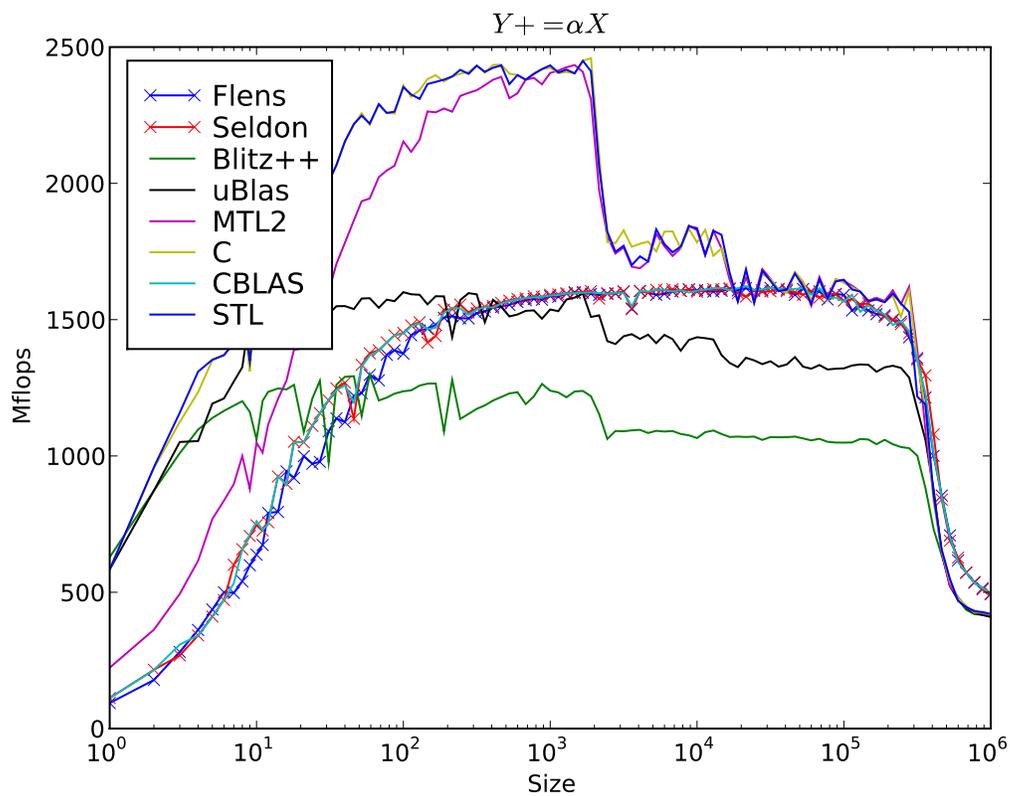}
  \caption{Y += alpha * X for dense vectors.}
  \label{fig:dense_Y+=alpha*X}
\end{figure}

\begin{figure}[htpb]
  \centering
  \includegraphics[width=\textwidth]{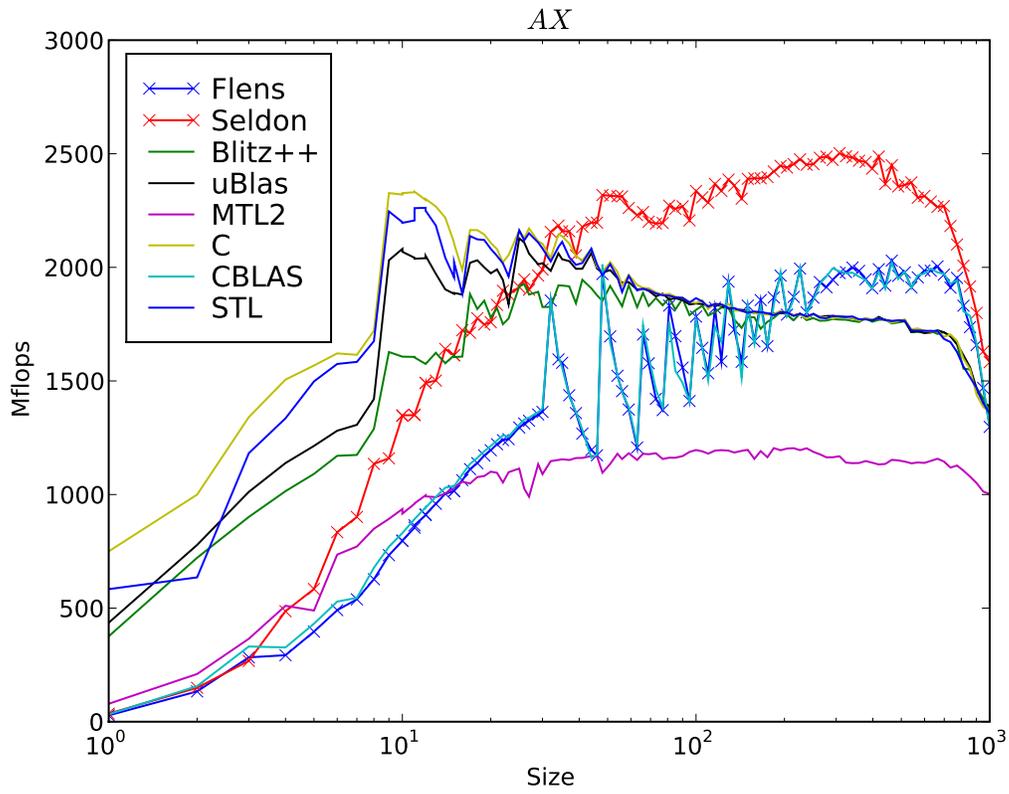}
  \caption{Matrix vector product for dense matrices and vectors.}
  \label{fig:dense_matrix_vector_product}
\end{figure}

\begin{figure}[htpb]
  \centering
  \includegraphics[width=\textwidth]{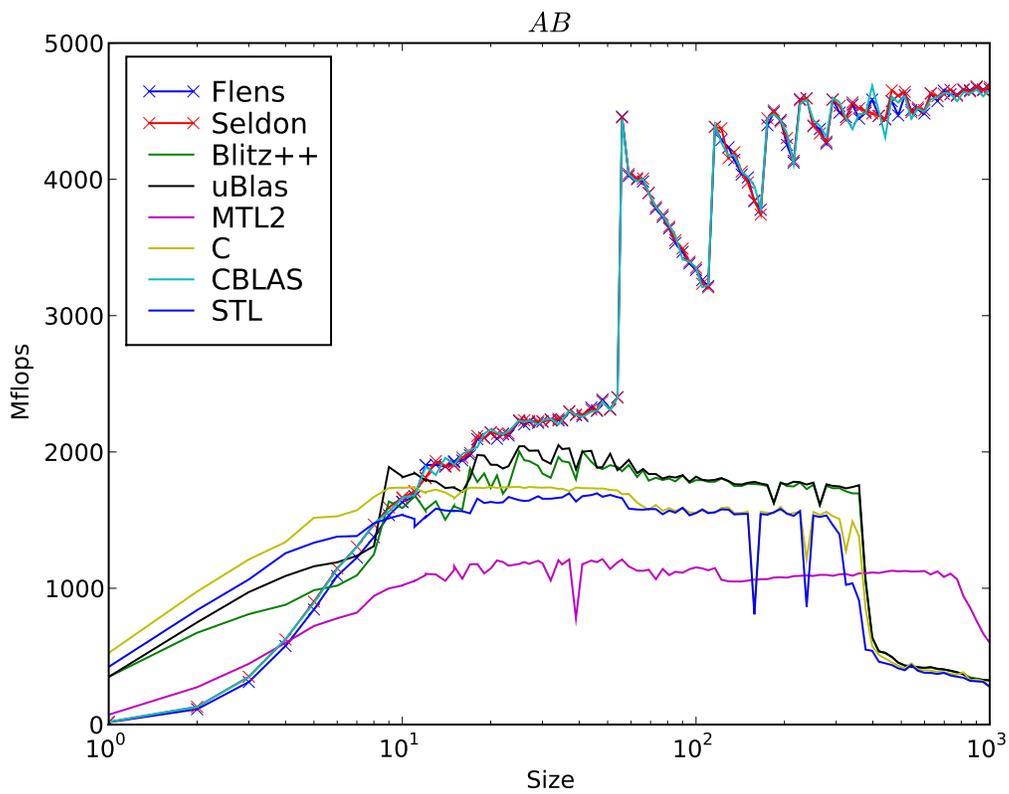}
  \caption{Matrix matrix product for dense matrices.}
  \label{fig:dense_matrix_matrix_product}
\end{figure}

\newpage
\subsubsection{Benchmarks for Sparse Matrix}

Adapted from \url{http://flens.sourceforge.net/session2/tut4.html}. \\
Compiled with -DNDEBUG -O3 options for g++.\\
Matrices of 1 000 000 lines.

\begin{enumerate}
\item Benchmark for sparse matrix, vector product with random initialization\\
5 non zero elements per line on average. In CRS Flens format, the number of
non zero values stored corresponds to the number of non zero values plus the
number of empty lines (one 0 is inserted on the diagonal).\\

\begin{tabular}{|l|c|c|}
\hline
& Flens & Seldon\\
\hline
initialization & 0.35s & \multirow{2}{*}{1.48s}\\
finalization & 1.23s &\\
\hline
y = Ax & 1.28s & 1.3s\\
y = A'x & 1.61s & 1.49s\\
\hline
\end{tabular}

\item{Benchmark with initialization in order and tridiagonal matrix}~\\
3 non zero elements per line (except on first and last lines with only 2 non
zero elements).\\

\begin{tabular}{|l|c|c|c|}
\hline
& Flens & Seldon  & Seldon \\
& & (matrix built by hand) & (matrix built with a generic algorithm *)\\
\hline
initialization & 0.28s & \multirow{2}{*}{0.08s} & \multirow{2}{*}{0.1s} \\
finalization & 0.16s & & \\
\hline
y = Ax & 0.19s & 0.18s & 0.18s \\
y = A'x & 0.22s & 0.22s & 0.22s\\
\hline
\end{tabular}

*using MergeSort on already sorted vectors, instead of QuickSort used in
random initialization case. \\

\item{Benchmark for computation of the residual}~\\
5 non zero elements per line on average.\\

\begin{tabular}{|l|c|c|}
\hline
& Flens & Seldon\\
\hline
initialization & 1.34s & \multirow{2}{*}{2.46s}\\
finalization & 1.25s &\\
\hline
computations r = b - Ax & 1.39s & 1.35s\\
\hline
\end{tabular}
\end{enumerate}

\newpage
\section*{Acknowledgement}
This document benefits from discussions with Vivien Mallet, Dominique Chapelle
and Philippe Moireau, and also from corrections thanks to Xavier Clerc.





\part{Appendix}

These sections deal with Flens distribution RC1 (Jul. 2007), Seldon
distribution 2008-11-12 and Trilinos distribution 9.0.1 (Oct. 2008). The first
two libraries satisfy the main requirements exposed in
section~\ref{part:requirements}, and Trilinos was supposed to be a good
reference as for the coverage of Blas and Lapack.

\section{Flens Overloaded Operator Performance Compared to Seldon}
\label{sec:flens_overloaded}
Flens implements a user-friendly interface to Blas routines by overloading
operators. For example, one can add vectors by using mathematical symbols:
\code{y = a + b + c}. Here are a few tests to check if this does not imply any
loss of performance in the computation.

In table \ref{table:flens_overloaded_operator_performance} are presented CPU
times measured with \code{std::clock()} for several operations, for
RowMajor (R.)  and ColMajor (C.) matrices. In the following, lower-case letters
denote vectors and upper-case letters denote matrices. \code{Y = A + B} cannot be
used for sparse matrices in Flens, so this has been tested only on general
dense matrices.

In ColumnMajor format for Flens, the affectation operation \code{Y = A} costs much
more (1.48s) than in RowMajor format (0.35s). This lack of performance slows
down sum operations such as \code{Y = A + B}. This is due to a direct call to Blas to
copy the data (a memory block copy in C++ is used in Seldon).

In ColumnMajor for Seldon, matrix vector product such as \code{Y += transpose(A)*B}
is three times longer than in Flens. This is due to a call to a generic
function instead of the Blas routine. Taking advantage of a local knowledge
of Seldon (one of its authors, Vivien Mallet, is part of my team!), we have
fixed this bug and got the same results as in Flens for \code{y~+=
  transpose(A)*b} and \code{y +=2*transpose(A)*b + 1.5*c + d}.

\begin{landscape}
  \begin{table}
    \caption{Flens and Seldon performance benchmarks using overloaded
      operators.}
    \label{table:flens_overloaded_operator_performance}
    \centering 
    \begin{tabular}{|l||c|c|c||c|c|c|}
      \hline
      Operation & Flens code & R.(s)  & C.(s)  & Seldon code & R.(s) & C.(s) \\
      \hline
      y = a + b + c & y = a + b + c; & 1.16 & 1.17 & y = a; Add(b,y); Add(c,y); &
      1.13  & 1.12 \\
      & y = a; y += b; y += c; & 1.16 & 1.18 & & & \\  
      & y = a + b; y += c; & 1.17 & 1.18 & & & \\
      \hline
      y = a + b + c + d & y=a+b+c+d; & 1.58 & 1.59 & y=a; Add(b,y); Add(c,y);
      Add(d,y) & 1.52 & 1.54\\
      & y = a; y += b; y += c; y += d; & 1.6 & 1.58 & & & \\
      \hline
      y += 4a & y += 4a; & 4.11 & 4.07 & Add(4.,a,y); & 4.09 & 4.09\\
      &  axpy(4, a, y); & 4.09 & 4.1 & & &\\
      \hline
      Y = A & Y = A; & 0.35 & 1.48 & Y = A; & 0.32 & 0.32 \\
      \hline
      Y = A + B & Y = A + B;  & 0.78 & 2.92 & Y = A; Add(B,Y); & 0.76 & 1.89\\
      &  Y = A; Y += B; & 0.78 & 2.93 & & & \\
      \hline
      Y = A + B + C & Y = A + B + C; & 1.21 & 4.41 & Y = A; Add(B,Y);
      Add(C,Y); & 1.24 & 3.44\\
      & Y = A + B; Y += C; & 1.22 & 4.35 & & & \\
      & Y = A; Y += B; Y += C; & 1.21 & 4.35 & & & \\
      \hline
      Y = A + B + C + D & Y = A + B + C + D; & 1.63 & 5.78 & Y = A; Add(B,Y); Add(C,Y);
      Add(D,Y); & 1.7 & 4.95\\
      & Y = A; Y+=B; Y+=C; Y+=D; & 1.62 & 5.78 & & & \\
      \hline
      y = Ax + b & y = Ax + b; & 1.64 & 2.09 & Copy(b,y); MltAdd(1.,A,x,1.,y); & 1.61 & 2.08\\
      \hline 
      y += A*b & y += A*b; & 1.63 & 2.07 & MltAdd(1.,A,b,1.,y); & 1.63 & 2.08\\
      \hline
      y += transpose(A)*b & y += transpose(A)*b; & 2.05 & 1.63 &
      MltAdd(1.,SeldonTrans,A,b,1.,y); & 6.64/2.05* & 1.53 \\
      \hline
      y+=2transpose(A)*b+1.5c+d & y += 2*transpose(A)*b+1.5c+d; & 2.06 &
      1.64 & MltAdd(2.,SeldonTrans,A,b,1.,y);Add(1.5,c,y);Add(d,y); & 6.66/2.05* &
      1.52  \\
      \hline
      C += 1.5A*transpose(B) & C += 1.5A*transpose(B); & 43.18 & 43.3 &
      MltAdd(1.5,SeldonNoTrans, A,SeldonTrans,B,1.,C); & 43.1 & 43.26\\
      \hline
    \end{tabular} 
  \end{table}
  * with the correction (see above for the explanation).
\end{landscape}

\section{Flens, Seldon and Trilinos Content Comparisons}
\label{sec:flens_seldon_trilinos_comparisons}

\subsection{Available Matrix Types from Blas (Flens and Seldon)}
~

F stands for Flens, S stands for Seldon. A black cell for an existing
structure; a gray cell for no structure.
\\

\begin{tabular}{|lcc|lcc|lcc|}
  \hline
  & F & S &  & F & S &  & F & S \\
  \hline
  GE - General & \True & \True & GB - General Band & \True & \False & & & \\
  SY - SYmmetric & \True & \True & SB - Sym. Band & \True & \False & SP -
  Sym. Packed & \True & \True \\
  HE - HErmitian & \False & \True &  HB - Herm. Band & \False & \False & HP -
  Herm. Packed & \False & \True \\
  TR - TRiangular & \True & \True & TB - Triang. Band & \True & \False & TP -
  Triang. Packed & \True & \True \\
  \hline
\end{tabular}

\newpage
\subsection{Available Interfaces to Blas and Lapack Routines (Flens and  Seldon)}
~

A black cell for an existing interface to a given routine; a gray cell for no
interface. 

\begin{newMargin}{-2cm} {-1cm}

\begin{enumerate}
\item{Blas routines}

\begin{tabular}{|lcc|lcc|lcc|lcc|}
  \hline
  & Flens & Seldon &  & Flens & Seldon &  & Flens & Seldon &  & Flens & Seldon
  \\
  \hline
  srotg & \False & \True & drotg & \False & \True & crotg & \False &
  \False & zrotg & \False & \False \\
  srotmg & \False & \True & drotmg & \False & \True & srot & \False &
  \True & drot & \False & \True \\
  csrot & \False & \False & zdrot & \False & \False & srotm & \False & \True & drotm & \False & \True \\
  sswap & \False & \True & dswap & \False & \True & cswap & \False & \True & zswap & \False & \True \\
  sscal & \True & \True & dscal & \True & \True & cscal & \False & \True & zscal & \True & \True \\
  csscal & \False & \True & zdscal & \False & \True & scopy & \True & \True & dcopy & \True & \True \\
  ccopy & \False & \True & zcopy & \True & \True & saxpy & \True & \True & daxpy & \True & \True \\
  caxpy & \False & \True & zaxpy & \True & \True & sdot & \True & \True & ddot & \True & \True \\
  sdsdot & \False & \True & dsdot & \False & \True & cdotu & \False & \True & zdotu & \False & \True \\
  cdotc & \False & \True & zdotc & \False & \True & snrm2 & \True & \True & dnrm2 & \True & \True \\
  scnrm2 & \False & \True & dznrm2 & \True & \True & sasum & \False & \True & dasum & \False & \True \\
  scasum & \False & \True & dzasum & \False & \True & isamax & \True & \True & idamax & \True & \True \\
  icamax & \False & \True & izamax & \True & \True & sgemv & \True & \True & dgemv & \True & \True \\
  cgemv & \False & \True & zgemv & \True & \True & sgbmv & \True & \False & dgbmv & \True & \False \\
  cgbmv & \False & \False & zgbmv & \False & \False & chemv & \False & \True & zhemv & \False & \True \\
  chbmv & \False & \False & zhbmv & \False & \False & chpmv & \False & \True & zhpmv & \False & \True \\
  ssymv & \True & \True & dsymv & \True & \True & ssbmv & \True & \False & dsbmv & \True & \False \\
  sspmv & \True & \True & dspmv & \True & \True & strmv & \True & \True & dtrmv & \True & \True \\
  ctrmv & \False & \True & ztrmv & \False & \True & stbmv & \True & \False & dtbmv & \True & \False \\
  ctbmv & \False & \False & ztbmv & \False & \False & stpmv & \True & \True & dtpmv & \True & \True \\
  ctpmv & \False & \True & ztpmv & \False & \True & strsv & \False & \True & dtrsv & \True & \True \\
  ctrsv & \False & \True & ztrsv & \False & \True & stbsv & \False & \False & dtbsv & \False & \False \\
  ctbsv & \False & \False & ztbsv & \False & \False & stpsv & \False & \True & dtpsv & \False & \True \\
  ctpsv & \False & \True & ztpsv & \False & \True & sger & \False & \True & dger & \False & \True \\
  cgeru & \False & \True & zgeru & \False & \True & cgerc & \False & \True & zgerc & \False & \True \\
  cher & \False & \True & zher & \False & \True & chpr & \False & \True & zhpr & \False & \True \\
  cher2 & \False & \False & zher2 & \False & \False & chpr2 & \False & \True & zhpr2 & \False & \True \\
  ssyr & \False & \True & dsyr & \False & \True & sspr & \False & \True & dspr & \False & \True \\
  ssyr2 & \False & \False & dsyr2 & \False & \False & sspr2 & \False & \True & dspr2 & \False & \True \\
  sgemm & \True & \True & dgemm & \True & \True & cgemm & \False & \True & zgemm & \True & \True \\
  ssymm & \True & \True & dsymm & \True & \True & csymm & \False & \True & zsymm & \False & \True \\
  chemm & \False & \True & zhemm & \False & \True & ssyrk & \False & \False & dsyrk & \False & \False \\
  csyrk & \False & \False & zsyrk & \False & \False & cherk & \False & \False & zherk & \False & \False \\
  ssyr2k & \False & \False & dsyr2k & \False & \False & csyr2k & \False & \False & zsyr2k & \False & \False \\
  cher2k & \False & \False & zher2k & \False & \False & strmm & \True & \True & dtrmm & \True & \True \\
  ctrmm & \False & \True & ztrmm & \False & \True & strsm & \True & \True & dtrsm & \True & \True \\
  ctrsm & \False & \True & ztrsm & \False & \True & & & & & & \\
  \hline
\end{tabular}

\begin{tabular}{|p{4.05cm}|cc|}
  \hline
  Blas & Flens & Seldon \\
  \hline
  Total & 44 & 110 \\
  Coverage & 30\% & 75\% \\
  \hline
\end{tabular}

\newpage
\item{Single precision real Lapack routines}

\begin{tabular}{|lcc|lcc|lcc|lcc|}
  \hline
  & Flens & Seldon &  & Flens & Seldon &  & Flens & Seldon &  & Flens & Seldon \\
  \hline
  sgesv & \True & \True & sgbsv & \True & \False & sgtsv & \False & \False & sposv & \False & \False \\
  sppsv & \False & \False & spbsv & \False & \False & sptsv & \False & \False & ssysv & \False & \False \\
  sspsv & \False & \False & sgels & \True & \False & sgelsd & \False & \False & sgglse & \False & \False \\
  sggglm & \False & \False & ssyev & \False & \True & ssyevd & \False & \False & sspev & \False & \True \\
  sspevd & \False & \False & ssbev & \False & \False & ssbevd & \False & \False & sstev & \False & \False \\
  sstevd & \False & \False & sgees & \False & \False & sgeev & \True & \True & sgesvd & \True & \True \\
  sgesdd & \False & \False & ssygv & \False & \True & ssygvd & \False & \False & sspgv & \False & \True \\
  sspgvd & \False & \False & ssbgv & \False & \False & ssbgvd & \False & \False & sgegs & \False & \False \\
  sgges & \False & \False & sgegv & \False & \False & sggev & \False & \True & sggsvd & \False & \False \\
  sgesvx & \False & \False & sgbsvx & \False & \False & sgtsvx & \False & \False & sposvx & \False & \False \\
  sppsvx & \False & \False & spbsvx & \False & \False & sptsvx & \False & \False & ssysvx & \False & \False \\
  sspsvx & \False & \False & sgelsx & \False & \False & sgelsy & \False & \False & sgelss & \True & \False \\
  ssyevx & \False & \False & ssyevr & \False & \False & ssygvx & \False & \False & sspevx & \False & \False \\
  sspgvx & \False & \False & ssbevx & \False & \False & ssbgvx & \False & \False & sstevx & \False & \False \\
  sstevr & \False & \False & sgeesx & \False & \False & sggesx & \False & \False & sgeevx & \False & \False \\
  sggevx & \False & \False & sbdsdc & \False & \False & sbdsqr & \False & \False & sdisna & \False & \False \\
  sgbbrd & \False & \False & sgbcon & \False & \False & sgbequ & \False & \False & sgbrfs & \False & \False \\
  sgbtrf & \True & \False & sgbtrs & \True & \False & sgebak & \False & \False & sgebal & \False & \False \\
  sgebrd & \False & \False & sgecon & \False & \True & sgeequ & \False & \True & sgehrd & \False & \False \\
  sgelqf & \False & \True & sgeqlf & \False & \False & sgeqp3 & \False & \False & sgeqpf & \False & \False \\
  sgeqrf & \True & \True & sgerfs & \False & \True & sgerqf & \False & \False & sgetrf & \True & \True \\
  sgetri & \True & \True & sgetrs & \True & \True & sggbak & \False & \False & sggbal & \False & \False \\
  sgghrd & \False & \False & sggqrf & \False & \False & sggrqf & \False & \False & sggsvp & \False & \False \\
  sgtcon & \False & \False & sgtrfs & \False & \False & sgttrf & \False & \False & sgttrs & \False & \False \\
  shgeqz & \False & \False & shsein & \False & \False & shseqr & \False & \False & sopgtr & \False & \False \\
  sopmtr & \False & \False & sorgbr & \False & \False & sorghr & \False & \False & sorglq & \False & \False \\
  sorgql & \False & \False & sorgqr & \True & \False & sorgrq & \False & \False & sorgtr & \False & \False \\
  sormbr & \False & \False & sormhr & \False & \False & sormlq & \False & \True & sormql & \False & \False \\
  sormqr & \True & \True & sormr3 & \False & \False & sormrq & \False & \False & sormrz & \False & \False \\
  sormtr & \False & \False & spbcon & \False & \False & spbequ & \False & \False & spbrfs & \False & \False \\
  spbstf & \False & \False & spbtrf & \False & \False & spbtrs & \False & \False & spocon & \False & \False \\
  spoequ & \False & \False & sporfs & \False & \False & spotrf & \False & \False & spotri & \False & \False \\
  spotrs & \False & \False & sppcon & \False & \False & sppequ & \False & \False & spprfs & \False & \False \\
  spptrf & \False & \False & spptri & \False & \False & spptrs & \False & \False & sptcon & \False & \False \\
  spteqr & \False & \False & sptrfs & \False & \False & spttrf & \False & \False & spttrs & \False & \False \\
  ssbgst & \False & \False & ssbtrd & \False & \False & sspcon & \False & \True & sspgst & \False & \False \\
  ssprfs & \False & \True & ssptrd & \False & \False & ssptrf & \False & \True & ssptri & \False & \True \\
  ssptrs & \False & \True & sstebz & \False & \False & sstedc & \False & \False & sstegr & \False & \False \\
  sstein & \False & \False & ssteqr & \False & \False & ssterf & \False & \False & ssycon & \False & \True \\
  ssygst & \False & \False & ssyrfs & \False & \True & ssytrd & \False & \False & ssytrf & \False & \True \\
  ssytri & \False & \True & ssytrs & \False & \True & stbcon & \False & \False & stbrfs & \False & \False \\
  stbtrs & \False & \False & stgevc & \False & \False & stgexc & \False & \False & stgsen & \False & \False \\
  stgsja & \False & \False & stgsna & \False & \False & stgsyl & \False & \False & stpcon & \False & \True \\
  stprfs & \False & \True & stptri & \False & \True & stptrs & \False & \True & strcon & \False & \True \\
  strevc & \False & \False & strexc & \False & \False & strrfs & \False & \True & strsen & \False & \False \\
  strsna & \False & \False & strsyl & \False & \False & strtri & \False & \True & strtrs & \True & \True \\
  stzrqf & \False & \False & stzrzf & \False & \False & & & & & & \\
  \hline
\end{tabular}

\begin{tabular}{|p{4.05cm}|cc|}
  \hline
  Single real Lapack & Flens & Seldon \\
  \hline
  Total & 15 & 36 \\
  Coverage & 8\% & 19\% \\
  \hline
\end{tabular}

\newpage
\item{Double precision real Lapack routines}

\begin{tabular}{|lcc|lcc|lcc|lcc|}
  \hline
  & Flens & Seldon &  & Flens & Seldon &  & Flens & Seldon &  & Flens & Seldon \\
  \hline
  dgesv & \True & \True & dsgesv & \False & \False & dgbsv & \True & \False & dgtsv & \False & \False \\
  dposv & \False & \False & dppsv & \False & \False & dpbsv & \False & \False & dptsv & \False & \False \\
  dsysv & \False & \False & dspsv & \False & \False & dgels & \True & \False & dgelsd & \False & \False \\
  dgglse & \False & \False & dggglm & \False & \False & dsyev & \False & \True & dsyevd & \False & \False \\
  dspev & \False & \True & dspevd & \False & \False & dsbev & \False & \False & dsbevd & \False & \False \\
  dstev & \False & \False & dstevd & \False & \False & dgees & \False & \False & dgeev & \True & \True \\
  dgesvd & \True & \True & dgesdd & \False & \False & dsygv & \False & \True & dsygvd & \False & \False \\
  dspgv & \False & \True & dspgvd & \False & \False & dsbgv & \False & \False & dsbgvd & \False & \False \\
  dgegs & \False & \False & dgges & \False & \False & dgegv & \False & \False & dggev & \False & \True \\
  dggsvd & \False & \False & dgesvx & \False & \False & dgbsvx & \False & \False & dgtsvx & \False & \False \\
  dposvx & \False & \False & dppsvx & \False & \False & dpbsvx & \False & \False & dptsvx & \False & \False \\
  dsysvx & \False & \False & dspsvx & \False & \False & dgelsx & \False & \False & dgelsy & \False & \False \\
  dgelss & \True & \False & dsyevx & \False & \False & dsyevr & \False & \False & dsygvx & \False & \False \\
  dspevx & \False & \False & dspgvx & \False & \False & dsbevx & \False & \False & dsbgvx & \False & \False \\
  dstevx & \False & \False & dstevr & \False & \False & dgeesx & \False & \False & dggesx & \False & \False \\
  dgeevx & \False & \False & dggevx & \False & \False & dbdsdc & \False & \False & dbdsqr & \False & \False \\
  ddisna & \False & \False & dgbbrd & \False & \False & dgbcon & \False & \False & dgbequ & \False & \False \\
  dgbrfs & \False & \False & dgbtrf & \True & \False & dgbtrs & \True & \False & dgebak & \False & \False \\
  dgebal & \False & \False & dgebrd & \False & \False & dgecon & \False & \True & dgeequ & \False & \True \\
  dgehrd & \False & \False & dgelqf & \False & \True & dgeqlf & \False & \False & dgeqp3 & \False & \True \\
  dgeqpf & \False & \False & dgeqrf & \True & \True & dgerfs & \False & \True & dgerqf & \False & \False \\
  dgetrf & \True & \True & dgetri & \True & \True & dgetrs & \True & \True & dggbak & \False & \False \\
  dggbal & \False & \False & dgghrd & \False & \False & dggqrf & \False & \False & dggrqf & \False & \False \\
  dggsvp & \False & \False & dgtcon & \False & \False & dgtrfs & \False & \False & dgttrf & \False & \False \\
  dgttrs & \False & \False & dhgeqz & \False & \False & dhsein & \False & \False & dhseqr & \False & \False \\
  dopgtr & \False & \False & dopmtr & \False & \False & dorgbr & \False & \False & dorghr & \False & \False \\
  dorglq & \False & \False & dorgql & \False & \False & dorgqr & \True & \True & dorgrq & \False & \False \\
  dorgtr & \False & \False & dormbr & \False & \False & dormhr & \False & \False & dormlq & \False & \True \\
  dormql & \False & \False & dormqr & \True & \True & dormr3 & \False & \False & dormrq & \False & \False \\
  dormrz & \False & \False & dormtr & \False & \False & dpbcon & \False & \False & dpbequ & \False & \False \\
  dpbrfs & \False & \False & dpbstf & \False & \False & dpbtrf & \False & \False & dpbtrs & \False & \False \\
  dpocon & \False & \False & dpoequ & \False & \False & dporfs & \False & \False & dpotrf & \False & \False \\
  dpotri & \False & \False & dpotrs & \False & \False & dppcon & \False & \False & dppequ & \False & \False \\
  dpprfs & \False & \False & dpptrf & \False & \False & dpptri & \False & \False & dpptrs & \False & \False \\
  dptcon & \False & \False & dpteqr & \False & \False & dptrfs & \False & \False & dpttrf & \False & \False \\
  dpttrs & \False & \False & dsbgst & \False & \False & dsbtrd & \False & \False & dspcon & \False & \True \\
  dspgst & \False & \False & dsprfs & \False & \True & dsptrd & \False & \False & dsptrf & \False & \True \\
  dsptri & \False & \True & dsptrs & \False & \True & dstebz & \False & \False & dstedc & \False & \False \\
  dstegr & \False & \False & dstein & \False & \False & dsteqr & \False & \False & dsterf & \False & \False \\
  dsycon & \False & \True & dsygst & \False & \False & dsyrfs & \False & \True & dsytrd & \False & \False \\
  dsytrf & \False & \True & dsytri & \False & \True & dsytrs & \False & \True & dtbcon & \False & \False \\
  dtbrfs & \False & \False & dtbtrs & \False & \False & dtgevc & \False & \False & dtgexc & \False & \False \\
  dtgsen & \False & \False & dtgsja & \False & \False & dtgsna & \False & \False & dtgsyl & \False & \False \\
  dtpcon & \False & \True & dtprfs & \False & \True & dtptri & \False & \True & dtptrs & \False & \True \\
  dtrcon & \False & \True & dtrevc & \False & \False & dtrexc & \False & \False & dtrrfs & \False & \True \\
  dtrsen & \False & \False & dtrsna & \False & \False & dtrsyl & \False & \False & dtrtri & \False & \True \\
  dtrtrs & \True & \True & dtzrqf & \False & \False & dtzrzf & \False & \False & & & \\
  \hline
\end{tabular}

\begin{tabular}{|p{4.05cm}|cc|}
  \hline
  Double real Lapack & Flens & Seldon \\
  \hline
  Total & 15 & 38 \\
  Coverage & 8\% & 20\% \\
  \hline
\end{tabular}

\newpage
\item{Single precision complex Lapack routines}

\begin{tabular}{|lcc|lcc|lcc|lcc|}
  \hline
  & Flens & Seldon &  & Flens & Seldon &  & Flens & Seldon &  & Flens & Seldon \\
  \hline
  cgesv & \False & \True & cgbsv & \False & \False & cgtsv & \False & \False & cposv & \False & \False \\
  cppsv & \False & \False & cpbsv & \False & \False & cptsv & \False & \False & csysv & \False & \False \\
  chesv & \False & \False & cspsv & \False & \False & chpsv & \False & \False & cgels & \False & \False \\
  cgelsd & \False & \False & cgglse & \False & \False & cggglm & \False & \False & cheev & \False & \True \\
  cheevd & \False & \False & cheevr & \False & \False & chpev & \False & \True & chpevd & \False & \False \\
  chbev & \False & \False & chbevd & \False & \False & cgees & \False & \False & cgeev & \True & \True \\
  cgesvd & \False & \True & cgesdd & \False & \False & chegv & \False & \True & chegvd & \False & \False \\
  chpgv & \False & \True & chpgvd & \False & \False & chbgv & \False & \False & chbgvd & \False & \False \\
  cgegs & \False & \False & cgges & \False & \False & cgegv & \False & \False & cggev & \False & \True \\
  cggsvd & \False & \False & cgesvx & \False & \False & cgbsvx & \False & \False & cgtsvx & \False & \False \\
  cposvx & \False & \False & cppsvx & \False & \False & cpbsvx & \False & \False & cptsvx & \False & \False \\
  csysvx & \False & \False & chesvx & \False & \False & cspsvx & \False & \False & chpsvx & \False & \False \\
  cgelsx & \False & \False & cgelsy & \False & \False & cgelss & \False & \False & cheevx & \False & \False \\
  cheevr & \False & \False & chegvx & \False & \False & chpevx & \False & \False & chpgvx & \False & \False \\
  chbevx & \False & \False & chbgvx & \False & \False & cgeesx & \False & \False & cggesx & \False & \False \\
  cgeevx & \False & \False & cggevx & \False & \False & cbdsdc & \False & \False & cbdsqr & \False & \False \\
  cgbbrd & \False & \False & cgbcon & \False & \False & cgbequ & \False & \False & cgbrfs & \False & \False \\
  cgbtrf & \False & \False & cgbtrs & \False & \False & cgebak & \False & \False & cgebal & \False & \False \\
  cgebrd & \False & \False & cgecon & \False & \True & cgeequ & \False & \True & cgehrd & \False & \False \\
  cgelqf & \False & \False & cgeqlf & \False & \False & cgeqp3 & \False & \False & cgeqpf & \False & \False \\
  cgeqrf & \False & \False & cgerfs & \False & \True & cgerqf & \False & \False & cgetrf & \False & \True \\
  cgetri & \False & \True & cgetrs & \False & \True & cggbak & \False & \False & cggbal & \False & \False \\
  cgghrd & \False & \False & cggqrf & \False & \False & cggrqf & \False & \False & cggsvp & \False & \False \\
  cgtcon & \False & \False & cgtrfs & \False & \False & cgttrf & \False & \False & cgttrs & \False & \False \\
  chgeqz & \False & \False & chsein & \False & \False & chseqr & \False & \False & cupgtr & \False & \False \\
  cupmtr & \False & \False & cungbr & \False & \False & cunghr & \False & \False & cunglq & \False & \False \\
  cungql & \False & \False & cungqr & \False & \False & cungrq & \False & \False & cungtr & \False & \False \\
  cunmbr & \False & \False & cunmhr & \False & \False & cunmlq & \False & \False & cunmql & \False & \False \\
  cunmqr & \False & \False & cunmr3 & \False & \False & cunmrq & \False & \False & cunmrz & \False & \False \\
  cunmtr & \False & \False & cpbcon & \False & \False & cpbequ & \False & \False & cpbrfs & \False & \False \\
  cpbstf & \False & \False & cpbtrf & \False & \False & cpbtrs & \False & \False & cpocon & \False & \False \\
  cpoequ & \False & \False & cporfs & \False & \False & cpotrf & \False & \False & cpotri & \False & \False \\
  cpotrs & \False & \False & cppcon & \False & \False & cppequ & \False & \False & cpprfs & \False & \False \\
  cpptrf & \False & \False & cpptri & \False & \False & cpptrs & \False & \False & cptcon & \False & \False \\
  cpteqr & \False & \False & cptrfs & \False & \False & cpttrf & \False & \False & cpttrs & \False & \False \\
  chbgst & \False & \False & chbtrd & \False & \False & cspcon & \False & \True & chpcon & \False & \True \\
  chpgst & \False & \False & csprfs & \False & \True & chprfs & \False & \True & chptrd & \False & \False \\
  csptrf & \False & \True & chptrf & \False & \True & csptri & \False & \True & chptri & \False & \True \\
  csptrs & \False & \True & chptrs & \False & \True & cstedc & \False & \False & cstegr & \False & \False \\
  cstein & \False & \False & csteqr & \False & \False & csycon & \False & \True & checon & \False & \True \\
  chegst & \False & \False & csyrfs & \False & \True & cherfs & \False & \True & chetrd & \False & \False \\
  csytrf & \False & \True & chetrf & \False & \True & csytri & \False & \True & chetri & \False & \True \\
  csytrs & \False & \True & chetrs & \False & \True & ctbcon & \False & \False & ctbrfs & \False & \False \\
  ctbtrs & \False & \False & ctgevc & \False & \False & ctgexc & \False & \False & ctgsen & \False & \False \\
  ctgsja & \False & \False & ctgsna & \False & \False & ctgsyl & \False & \False & ctpcon & \False & \True \\
  ctprfs & \False & \True & ctptri & \False & \True & ctptrs & \False & \True & ctrcon & \False & \True \\
  ctrevc & \False & \False & ctrexc & \False & \False & ctrrfs & \False & \True & ctrsen & \False & \False \\
  ctrsna & \False & \False & ctrsyl & \False & \False & ctrtri & \False & \True & ctrtrs & \False & \True \\
  ctzrqf & \False & \False & ctzrzf & \False & \False & & & & & & \\
  \hline
\end{tabular}

\begin{tabular}{|p{4.05cm}|cc|}
  \hline
  Single complex Lapack & Flens & Seldon \\
  \hline
  Total & 1 & 42 \\
  Coverage & 1\% & 22\% \\
  \hline
\end{tabular}

\newpage
\item{Double precision complex Lapack routines}

\begin{tabular}{|lcc|lcc|lcc|lcc|}
  \hline
  & Flens & Seldon &  & Flens & Seldon &  & Flens & Seldon &  & Flens & Seldon \\
  \hline
  zgesv & \False & \True & zgbsv & \False & \False & zgtsv & \False & \False & zposv & \False & \False \\
  zppsv & \False & \False & zpbsv & \False & \False & zptsv & \False & \False & zsysv & \False & \False \\
  zhesv & \False & \False & zspsv & \False & \False & zhpsv & \False & \False & zgels & \True & \False \\
  zgelsd & \False & \False & zgglse & \False & \False & zggglm & \False & \False & zheev & \False & \True \\
  zheevd & \False & \False & zhpev & \False & \True & zhpevd & \False & \False & zhbev & \False & \False \\
  zhbevd & \False & \False & zgees & \False & \False & zgeev & \True & \True & zgesvd & \False & \True \\
  zgesdd & \False & \False & zhegv & \False & \True & zhegvd & \False & \False & zhpgv & \False & \True \\
  zhpgvd & \False & \False & zhbgv & \False & \False & zhbgvd & \False & \False & zgegs & \False & \False \\
  zgges & \False & \False & zgegv & \False & \False & zggev & \False & \True & zggsvd & \False & \False \\
  zgesvx & \False & \False & zgbsvx & \False & \False & zgtsvx & \False & \False & zposvx & \False & \False \\
  zppsvx & \False & \False & zpbsvx & \False & \False & zptsvx & \False & \False & zsysvx & \False & \False \\
  zhesvx & \False & \False & zspsvx & \False & \False & zhpsvx & \False & \False & zgelsx & \False & \False \\
  zgelsy & \False & \False & zgelss & \True & \False & zheevx & \False & \False & zheevr & \False & \False \\
  zhegvx & \False & \False & zhpevx & \False & \False & zhpgvx & \False & \False & zhbevx & \False & \False \\
  zhbgvx & \False & \False & zgeesx & \False & \False & zggesx & \False & \False & zgeevx & \False & \False \\
  zggevx & \False & \False & zbdsdc & \False & \False & zbdsqr & \False & \False & zgbbrd & \False & \False \\
  zgbcon & \False & \False & zgbequ & \False & \False & zgbrfs & \False & \False & zgbtrf & \False & \False \\
  zgbtrs & \False & \False & zgebak & \False & \False & zgebal & \False & \False & zgebrd & \False & \False \\
  zgecon & \False & \True & zgeequ & \False & \True & zgehrd & \False & \False & zgelqf & \False & \True \\
  zgeqlf & \False & \False & zgeqp3 & \False & \False & zgeqpf & \False & \False & zgeqrf & \False & \True \\
  zgerfs & \False & \True & zgerqf & \False & \False & zgetrf & \True & \True & zgetri & \True & \True \\
  zgetrs & \False & \True & zggbak & \False & \False & zggbal & \False & \False & zgghrd & \False & \True \\
  zggqrf & \False & \False & zggrqf & \False & \False & zggsvp & \False & \False & zgtcon & \False & \False \\
  zgtrfs & \False & \False & zgttrf & \False & \False & zgttrs & \False & \False & zhgeqz & \False & \True \\
  zhsein & \False & \False & zhseqr & \False & \False & zupgtr & \False & \False & zupmtr & \False & \False \\
  zungbr & \False & \False & zunghr & \False & \False & zunglq & \False & \False & zungql & \False & \False \\
  zungqr & \False & \True & zungrq & \False & \False & zungtr & \False & \False & zunmbr & \False & \False \\
  zunmhr & \False & \False & zunmlq & \False & \True & zunmql & \False & \False & zunmqr & \False & \True \\
  zunmr3 & \False & \False & zunmrq & \False & \False & zunmrz & \False & \False & zunmtr & \False & \False \\
  zpbcon & \False & \False & zpbequ & \False & \False & zpbrfs & \False & \False & zpbstf & \False & \False \\
  zpbtrf & \False & \False & zpbtrs & \False & \False & zpocon & \False & \False & zpoequ & \False & \False \\
  zporfs & \False & \False & zpotrf & \False & \False & zpotri & \False & \False & zpotrs & \False & \False \\
  zppcon & \False & \False & zppequ & \False & \False & zpprfs & \False & \False & zpptrf & \False & \False \\
  zpptri & \False & \False & zpptrs & \False & \False & zptcon & \False & \False & zpteqr & \False & \False \\
  zptrfs & \False & \False & zpttrf & \False & \False & zpttrs & \False & \False & zhbgst & \False & \False \\
  zhbtrd & \False & \False & zspcon & \False & \True & zhpcon & \False & \True & zhpgst & \False & \False \\
  zsprfs & \False & \True & zhprfs & \False & \True & zhptrd & \False & \False & zsptrf & \False & \True \\
  zhptrf & \False & \True & zsptri & \False & \True & zhptri & \False & \True & zsptrs & \False & \True \\
  zhptrs & \False & \True & zstedc & \False & \False & zstegr & \False & \False & zstein & \False & \False \\
  zsteqr & \False & \False & zsycon & \False & \True & zhecon & \False & \True & zhegst & \False & \False \\
  zsyrfs & \False & \True & zherfs & \False & \True & zhetrd & \False & \False & zsytrf & \False & \True \\
  zhetrf & \False & \True & zsytri & \False & \True & zhetri & \False & \True & zsytrs & \False & \True \\
  zhetrs & \False & \True & ztbcon & \False & \False & ztbrfs & \False & \False & ztbtrs & \False & \False \\
  ztgevc & \False & \False & ztgexc & \False & \False & ztgsen & \False & \False & ztgsja & \False & \False \\
  ztgsna & \False & \False & ztgsyl & \False & \False & ztpcon & \False & \True & ztprfs & \False & \True \\
  ztptri & \False & \True & ztptrs & \False & \True & ztrcon & \False & \True & ztrevc & \False & \False \\
  ztrexc & \False & \False & ztrrfs & \False & \True & ztrsen & \False & \False & ztrsna & \False & \False \\
  ztrsyl & \False & \False & ztrtri & \False & \True & ztrtrs & \False & \True &
  ztzrqf & \False & \False \\
  ztzrzf & \False & \False & & & & & & & & & \\
  \hline
\end{tabular}

\begin{tabular}{|p{4.05cm}|cc|}
  \hline
  Double complex Lapack & Flens & Seldon \\
  \hline
  Total & 5 & 49 \\
  Coverage & 3\% & 25\% \\
  \hline
\end{tabular}
\end{enumerate}
\end{newMargin} 

\newpage
\subsection{Available Interfaces to Blas and Lapack Routines (Trilinos)}
~ 

As a rough guide, here are some results for Trilinos. Several Trilinos
packages offer at least a partial interface to Blas and Lapack routines:
\textit{Epetra}, \textit{Teuchos}, \textit{Amesos}, \textit{AztecOO},
\textit{ML}, \textit{MOOCHO} and \textit{Pliris}. The Trilinos column refers
to all the packages tested together. This result is the maximum coverage and
one should be careful for its interpretation. Indeed, some packages may not
communicate together, with non compatible structures, and therefore could not
be used together. Moreover, some of the interfaces may not be usable directly,
only indirectly through other functions. \\

\begin{newMargin}{-2cm} {-1cm}

\begin{tabular}{|p{4.05cm}|cccccccc|}
  \hline
  Blas & Trilinos &
  \textit{Epetra} & \textit{Teuchos} & \textit{Amesos} & \textit{AztecOO} &
  \textit{ML} & \textit{MOOCHO} & \textit{Pliris} \\
  \hline
  Total & 87 & 28 & 62 & 18 & 21 & 29 & 34 & 28\\
  Coverage & 60\% & 19\% & 42\% & 12\%  & 14\% & 20\% & 23\% & 19\% \\
  \hline
\end{tabular}

\begin{tabular}{|p{4.05cm}|cccccccc|}
  \hline
  Single real Lapack & Trilinos &
  \textit{Epetra} & \textit{Teuchos} & \textit{Amesos} & \textit{AztecOO} &
  \textit{ML} & \textit{MOOCHO} & \textit{Pliris} \\
  \hline
  Total & 57 & 44 & 44 & 4 & 8 & 32 & 1 & 1\\
  Coverage & 31\% & 24\% & 24\% & 2\% & 4\% & 17\%  & 1\% & 1\% \\
  \hline
\end{tabular}

\begin{tabular}{|p{4.05cm}|cccccccc|}
  \hline
  Double real Lapack & Trilinos &
  \textit{Epetra} & \textit{Teuchos} & \textit{Amesos} & \textit{AztecOO} &
  \textit{ML} & \textit{MOOCHO} & \textit{Pliris} \\
  \hline
  Total & 59 & 43 & 44 & 4 & 7 & 32 & 7 & 0 \\
  Coverage & 32\% & 23\% & 24\% & 2\% & 4\% & 17\% & 4\% & 0\% \\
  \hline
\end{tabular}

\begin{tabular}{|p{4.05cm}|cccccccc|}
  \hline
Single complex Lapack & Trilinos &
  \textit{Epetra} & \textit{Teuchos} & \textit{Amesos} & \textit{AztecOO} &
  \textit{ML} & \textit{MOOCHO} & \textit{Pliris} \\
  \hline
  Total & 39 & 1 & 38 & 1 & 1 & 1 & 0 & 0 \\
  Coverage & 20\% & 1\% & 20\% & 1\% & 1\% & 1\%  & 0\% & 0\% \\
  \hline
\end{tabular}

\begin{tabular}{|p{4.05cm}|cccccccc|}
  \hline
  Double complex Lapack & Trilinos &
  \textit{Epetra} & \textit{Teuchos} & \textit{Amesos} & \textit{AztecOO} &
  \textit{ML} & \textit{MOOCHO} & \textit{Pliris} \\
  \hline
  Total & 38 & 0 & 38 & 1 & 0 & 0 & 0 & 0\\
  Coverage & 20\% & 0\% & 20\% & 1\%  & 0\%  & 0\% & 0\% & 0\% \\
  \hline
\end{tabular}

\end{newMargin} 
\newpage
\section{Flens and Seldon Synoptic Comparison}

\begin{newMargin}{-1.5cm} {-1cm}
  \begin{tabular}{|p{2.5cm}||p{2.5cm}|p{2.5cm}||p{2.5cm}|p{2.5cm}||p{1.5cm}|}
    \hline
    & \multicolumn{2}{c||}{Flens}  &  \multicolumn{2}{c||}{Seldon}  & See\\
    & ~~~~~~~~~--- & ~~~~~~~~~ + & ~~~~~~~~~ --- & ~~~~~~~~~ + & section \\
    \hline
    Portability & Not recently tested on Windows & & & & \\
    \hline
    High-level ~ ~ interface & & & & Python interface generated by SWIG & \\
    \hline
    C++ ~ ~ ~ ~ ~ templates & & Supported & & Supported & \\
    \hline
    Matrix types & No hermitian matrices & & No band matrices & & \\
    \hline
    Sparse ~ ~ ~ ~ matrices & & General and symmetric, compressed row storage & &
    Harwell-Boeing and array of sparse vectors & \\
    \hline
    Sparse vectors & Not Supported & & & Supported & \\
    \hline
    Syntax & & Natural mathematical notation & & & \\
    \hline
    Maintenance & Release candidate RC1 in Jul. 2007, last commit Jan. 2009 &
    & & Latest release in  Nov. 2008 & \\
    \hline
    Blas and Lapack interface & & & & Good coverage &
    \ref{sec:flens_seldon_trilinos_comparisons} \\
    \hline
    Performance & Costly affectation in ColumnMajor format & & One bug (fixed
    in later versions) & Better performance for dense matrix vector product &
    \ref{sec:benchmark_seldon} and \ref{sec:flens_overloaded}\\
    \hline
    Vector and matrix views & & Supported & Not supported & & \\
    \hline
    Technical mastery & & & & One author in my INRIA team! & \\
    \hline
    Eigenvalues computation & Not Supported & &  & Eigenvalues and
    eigenvectors computation &  \\
    \hline
  \end{tabular}
\end{newMargin}

\end{document}